\documentclass[superscriptaddress,twocolumn,showpacs,preprintnumbers,prb]{revtex4}
\usepackage{graphicx}
\usepackage{times}

\usepackage{color}

\begin{document}

\title{Single-Band Model for Diluted Magnetic Semiconductors: \\
Dynamical and Transport Properties and \\
Relevance of Clustered States}

\author{G. Alvarez}
\affiliation{National High Magnetic Field Lab and Department of Physics, 
Florida State University, Tallahassee, FL 32310}

\author{E. Dagotto}
\affiliation{National High Magnetic Field Lab and Department of Physics, 
Florida State University, Tallahassee, FL 32310}

\date{\today}

\begin{abstract}
Dynamical and transport properties of a simple single-band spin-fermion lattice
model for (III,Mn)V diluted magnetic 
semiconductors (DMS) is here discussed using Monte Carlo simulations.
This effort is a continuation of previous work
(G. Alvarez {\it et al.}, Phys. Rev. Lett. {\bf 89}, 277202 (2002)) 
where the static properties of the model were studied.
The present results support the view that the relevant regime of $J/t$ 
(standard notation) is that of intermediate coupling,  where 
carriers are only partially trapped near Mn spins, and locally ordered regions 
(clusters) are present above the Curie temperature $T_{\rm C}$. 
This conclusion is based on 
the calculation of the resistivity vs. temperature, that shows  
a soft metal to insulator transition near $T_{\rm C}$, as well on
the analysis of the density-of-states and optical
conductivity. In addition, in the clustered regime a 
large magnetoresistance is observed in simulations. 
Formal analogies between DMS and manganites are also discussed.

\end{abstract}

\maketitle

\section{Introduction}

Diluted magnetic semiconductors (DMS) are attracting 
much attention lately due to their potential for device applications in the
growing field of spintronics. In particular, a large number of 
DMS studies have focused on III-V compounds where Mn doping in InAs 
and GaAs has been achieved using molecular beam epitaxy (MBE) techniques.
The main result of recent experimental efforts 
is the discovery\cite{ohno1,ohno,katsumoto,potashnik}
of ferromagnetism at a Curie temperature $T_{\rm C}$$\sim$110~K in
Ga$_{1-x}$Mn$_x$As, with Mn concentrations $x$ up to 10\%. It is widely
believed that this ferromagnetism is ``carried induced'', with holes
donated by Mn ions mediating a ferromagnetic interaction between
the randomly localized Mn$^{2+}$ spins. In practice, 
anti-site defects reduce the number of holes $n$ from its ideal 
value $n$=$x$, leading to a ratio $p$=$(n/x)$ substantially smaller than 1. 

Until recently, theoretical descriptions of DMS materials 
could be roughly classified in two categories. On one hand, the multi-band
nature of the problem is emphasized as a crucial aspect to quantitatively
understand these materials.\cite{macdonald1,schliemann,sinova,dietl} In this context
the lattice does not
play a key role and a continuum formulation is sufficient. The influence 
of disorder is considered on average. On the other hand, formulations
based on the possible strong localization of carriers at the
Mn-spin sites have also been proposed.\cite{bhatt,paper0} In this context
a single impurity-band description is considered 
sufficient for these materials.
Still within the single-band framework, but with carrier hopping not 
restricted to the Mn locations,
recent approaches to the problem
have used dynamical mean-field\cite{chatto,hwang} 
or reduced-basis\cite{calderon} approximations.
An
effective Hamiltonian for Ga$_{1-x}$Mn$_x$As 
was derived in the
dilute limit and studied in Ref.~\onlinecite{zarand}. 
All these calculations are important in our collective effort to
understand DMS materials. 

However, it is desirable to obtain a more
general view of the problem of ferromagnetism induced by
a diluted set of spins and holes. To reach this goal it would be better
to use techniques that do not rely
on mean-field approximations and, in addition, select a model
that has both the continuum and impurity-band formulations as limiting
cases. Such an approach would provide information
on potential procedures to further enhance $T_{\rm C}$ and clarify the
role of the many parameters in the problem. In addition, these general
considerations will be useful beyond the specific details of
Ga$_{1-x}$Mn$_x$As, allowing us to reach conclusions for other DMS. 
An effort in this direction was recently initiated
by Mayr and two of the authors.\cite{paper1} A detailed Monte Carlo study of a simple
model for DMS already allowed us to argue that $T_{\rm C}$ could
be further enhanced --perhaps even to room temperature-- by further
increasing $x$ and $p$ from current values in DMS samples.
These predictions seem in agreement with recent experimental developments
since very recently 
Ga$_{1-x}$Mn$_x$As samples with $T_{\rm C}$ as high as $150$ K were 
prepared,\cite{ku} a result believed to be caused by an enhanced free-hole density. 
Also samples with $T_{\rm C}\sim127$ K and $x>0.08$ were reported in 
 Ref.~\onlinecite{kuryliszyn},
and, very recently, a $T_{\rm C}$ of 140K was achieved on high quality GaMnAs films 
grown with Arsenic dimers.\cite{edmonds}
There seems to be plenty of room to further 
increase the critical temperatures according to these theoretical calculations.

The model used in Ref.~\onlinecite{paper1}
was a single-band lattice Hamiltonian, also studied by other groups, which 
does not have the multi-band characteristics and spin-orbit couplings
of the real problem. However,  it contains spin and holes in interaction
and it is 
expected to capture the main qualitative aspects of carrier-induced 
ferromagnetism in DMS materials. 
The choice of a single-band model allows us to focus on the essential 
aspects of the problem,
leaving aside the numerical complexity 
of the multi-band Hamiltonian, which unfortunately cannot be studied  
with reliable numerical techniques at present without introducing further 
approximations.\cite{comment3} Note that a quite 
similar philosophy
has been followed in the related area of manganites for many years. In fact,
it is well-known that the orbital order present in those compounds
can only be studied with a two-band model. However, single-band
approaches are sufficient to investigate the competition between 
ferromagnetic and anti-ferromagnetic states, an effect recently argued to be
at the heart of the colossal magnetoresistance phenomenon.\cite{dagotto,book} 
As a consequence, a qualitatively reliable study of the single-band
DMS model -- with itinerant holes and localized classical spins --
is expected to be important for progress in the DMS context as well.
\begin{figure}
\centering{
\includegraphics[width=7cm]{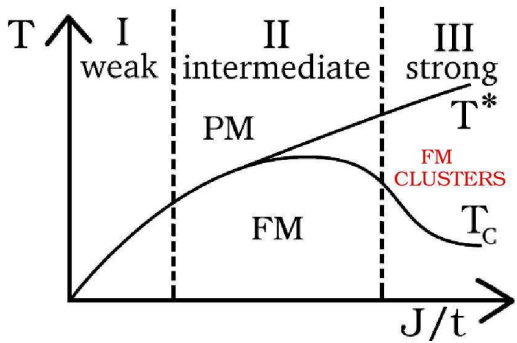}}
\caption{Phase diagram of the single-band model Eq.~(\ref{eq:ham}), as discussed 
in Ref.~\onlinecite{paper1}. The figure shows
schematically the three coupling regions discussed in the text: (I)
weak, (II) intermediate, and (III) strong coupling, as well as the $T_{\rm C}$ and $T^*$ 
dependence on the
coupling $J/t$. The region between $T^*$ and $T_{\rm C}$ 
contains FM ``clusters'', with magnetic moments that are not
aligned.\label{fig:summary}}
\end{figure}

In our previous publication\cite{paper1} the phase diagram of the
single-band model for DMS was already sketched.\cite{comment1}
This phase diagram is reproduced in Fig.\ref{fig:summary}.
The only parameter
in the Hamiltonian is the ratio $J/t$, where $J$ is the local 
anti-ferromagnetic coupling between the spins of carriers and localized
Mn$^{2+}$ ions, and $t$ is the carrier-hopping amplitude
(for details, see the full form of the Hamiltonian in the next section).
Due to its large value 5/2,
the Mn spin is assumed to be
classical for numerical simplicity.
At small $J/t$, individual carriers are only weakly bounded to the Mn spins.
Even for small realistic values of $x$, the carrier wave functions
strongly overlap and a large-bandwidth itinerant band dominates the physics.
In this regime, $T_{\rm C}$ grows with $J/t$. This portion 
of the phase diagram is referred to
as `weak coupling' in Fig.\ref{fig:summary}, and in our opinion 
it qualitatively corresponds to the continuum-limit approach pursued 
in Refs.~\onlinecite{macdonald1,schliemann,sinova}. In the other extreme of 
large $J/t$, the previous
effort\cite{paper1} found a $T_{\rm C}$ that is much
suppressed, converging to zero as $J/t$ increases 
(at least at small values of $x$). The reason for this behavior is the
strong localization of carriers at Mn spin locations, to take advantage
of the strong $J$ coupling. The localization
suppresses the mobility and the carrier-induced mechanism is no longer
operative. However, when several Mn spins are close to one another 
small regions can be magnetized efficiently. As a consequence, 
a picture emerges where small islands of ferromagnetism
are produced at a fairly large temperature scale $T^*$ ($>$$T_{\rm C}$)
that grows with $J$, 
but a global ferromagnetic state is only achieved upon further reducing
the temperature such that the overlaps of wave functions induce 
a percolation-like process that aligns the individual preformed clusters
The proper procedure to distinguish between these two regimes is through an
analysis of short- and long-distance spin correlations.
In our opinion, 
the effort of Refs.~\onlinecite{bhatt,paper0} belongs to this class,
and it corresponds to the `strong coupling' region of Fig.\ref{fig:summary}. 
Between the weak and strong coupling domains, an {\it intermediate region}
provides a natural interpolation between those 
two extreme cases. 
The Curie temperature is optimal (i.e., the largest) in this intermediate zone,
at fixed values of $p$ and $x$. In this regime, there is sufficient 
interaction between
the clusters to become globally ferromagnetic, and at the same time the $J$ is 
sufficiently strong to induce a robust $T_{\rm C}$. If chemical control
over $J/t$ were possible, this intermediate region would be the most
promising to increase the Curie temperature (again, at fixed $x$ and $p$). 

However, it may even occur
that the DMS materials are already in the intermediate optimal range of $J/t$ couplings.
An indirect way to test this hypothesis relies on calculations of
dynamical and transport properties, which are presented in 
this paper, and its comparison with experiments. 
In particular, the experimentally measured 
d.c. resistivity $\rho_{\rm dc}$ of DMS compounds
has a nontrivial shape with a (poor) metallic (d$\rho_{\rm dc}$/d$T$$>$0) behavior
below $T_{\rm C}$, which turns to insulating (d$\rho_{\rm dc}$/d$T$$<$0) at 
higher temperatures.
This nontrivial profile resembles results reported in
the area of manganites, which also have a metal-insulator transition as a function
of temperature,\cite{book} although in those materials the changes in resistivity
with temperature are far more dramatic than in DMS. The 
formal similarities DMS-manganites have already been remarked in
previous literature,\cite{paper1} and suggest a common origin of the
$\rho_{\rm dc}$ vs. temperature curves. In Mn-oxides it is believed that
above the Curie temperature, preformed ferromagnetic clusters with 
random orientations contribute to the insulating behavior 
of those materials.\cite{dagotto,book} For DMS, a similar rationalization
can be envisioned if the state of relevance above $T_{\rm C}$ 
has preformed magnetic clusters. From the previous discussion\cite{paper1}
it is known that the intermediate $J/t$
region can have clusters without collapsing $T_{\rm C}$ 
to a very small value. At the Curie temperature,
the alignment of these preformed moments leads to a metallic state.
An explicit calculation of the resistivity -- using 
techniques borrowed from the 
mesoscopic physics context -- is provided in this paper, and the results
support the conjecture that clustered states can explain the transport
properties of DMS.
These unveiled analogies with manganites are not just accidental. Clearly, 
approaches to DMS that rely on mobile carriers and localized spins in
interaction have close similarities with the standard single-band double-exchange model
where only one $e_g$-orbital is considered. The key difference 
is the presence of strong dilution in DMS as compared with manganites.

It is interesting to remark that 
in agreement with the `clustered' state described here,
recently Timm and von Oppen have shown that {\it correlated}
defects in DMS are needed
to describe experimental data.\cite{timm} Their simulations 
with Coulombic effects incorporated lead to cluster formation, with sizes well
beyond those obtained from a random distribution of Mn sites as considered here.
In this respect, the results of Ref.~\onlinecite{timm} provide an even more dramatic
clustered state than ours. If the
Mn spins were not distributed randomly in our simulations 
but in a correlated manner, 
the state above $T_{\rm C}$ would be even more insulating than reported 
below and the physics would resemble much closer that observed
in manganites. Note also that Kaminski and Das Sarma have also independently
arrived to a polaronic state\cite{kaminski} that qualitatively resembles
the clustered state discussed here.
It is also interesting that in recent ion-implanted (Ga,Mn)P:C experiments
that reported a high Curie temperature, the presence of 
ferromagnetic clusters were observed and they were attributed to disorder
and the proximity to a metal-insulator transition.\cite{hebard}
Even directly in (Ga,Mn)As, 
the inverse magnetic susceptibility deviates from the 
Curie law at temperatures above $T_{\rm C}$ (Fig.2(b) of
Ref.~\onlinecite{ohno}), which may be an indication
of an anomalous behavior. In manganites, these same anomalies in the
susceptibility are indeed taken as evidence
of the formation of clusters.\cite{dagotto}

The paper is organized as follows: In Section \ref{sec:method} the details of the
method of simulation, definitions, and conventions are presented. 
Section \ref{sec:dos} describes the results for the density-of-states (DOS) of
this model, the appearance of an ``impurity band'' at large $J/t$,
and the relation of the DOS with the optimal $T_{\rm C}$. 
The temperature and carrier dependence of
the optical conductivity and Drude weight are presented 
in Section \ref{sec:sigma}, with results compared with experiments. 
In Section \ref{sec:conductance}, the resistance of a small cluster is calculated
and also compared with experimental data. The intermediate coupling regime is the
only one found to qualitatively reproduce the available
resistivity vs. temperature curves. From this perspective, both the weak and strong 
coupling limits are not realistic to address DMS materials.
In Section \ref{sec:magnet}, the influence of a magnetic field 
is studied by calculating the magnetoresistance. Large effects are observed.
Section \ref{sec:cluster} describes the characteristics 
of the clustered-state regime above $T_{\rm C}$ at
strong coupling. Finally, in Section
\ref{sec:conclusion} the `transport' phase diagram is presented, and it is 
discussed in what region it appears to be the most
relevant to qualitatively reproduce results for Ga$_{1-x}$Mn$_x$As MBE-grown films.

\section{Method and Definitions} \label{sec:method}

\subsection{Hamiltonian Formalism}
There are two degrees of freedom in the DMS problem described here: 
(i) the local magnetic moments 
corresponding to the 5 electrons in the $d$ shell of each Mn
impurity, with a total spin 5/2, and (ii) the itinerant carriers
 produced by the Mn impurities.
Since these Mn impurities substitute Ga atoms in the zinc-blende 
structure of GaAs, there is, in principle, one hole per Mn. 
However, it has been found that the system is
heavily compensated and, as a consequence, the actual concentration 
of carriers is lower. In the present study, both the density of Mn atoms, $x$, 
and the density of carriers, $n$, are
treated as independent input parameters. 
Moreover, the ratio $p=n/x$ is defined, which is a measure of
the compensation of the system, e.g. for $p=0$ the system is totally 
compensated and for $p=1$ there is no compensation.\\
\indent The Hamiltonian of the system in the one-band approximation can be written as:
\begin{equation}
{\hat H}=-t\sum_{<ij>,\sigma}{\hat c}^\dagger_{i\sigma} {\hat c}_{j\sigma} +
J\sum_{I}\vec{S}_I\cdot\vec{\sigma}_I,
\label{eq:ham}
\end{equation}
\noindent where 
${\hat c}^\dagger_{i\sigma}$ creates a carrier at site $i$
with spin $\sigma$. The carrier-spin operator interacting
antiferromagnetically with the localized Mn-spin $\vec{S}_I$ is
$\vec{\sigma}_I={\hat
c}^\dagger_{I\alpha}\vec{\sigma}_{\alpha,\beta}{\hat c}_{I\beta}$. 
Through nearest-neighbor hopping, 
the carriers can hop to $any$ site of the square or cubic lattice. 
The interaction term is restricted to a randomly selected but fixed 
set of sites, denoted by $I$. Note that in addition to the use of
a single band, there are other approximation in the model described
here: (i) The Hubbard $U/t$ is not included. This is justified based
on the low-carrier concentration of the problem, since in this case
the probability of double occupancy is small. In addition, Mn-oxide
investigations\cite{dagotto,book} have shown that an intermediate or
large $J$ coupling acts similarly as $U/t$, also suppressing double
occupancy. (ii) A nearest-neighbors 
anti-ferromagnetic coupling between the Mn spins is
not included. Again, this is justified as long as $x$ is small. 
(iii) Finally, also potential disorder is neglected (together
with the spin, the Mn sites should in principle act 
as charge trapping centers).
This approximation is not problematic when the Fermi energy is larger than the width of
the disorder potential (i.e. large $p$ and $x$). However, in the ($p$, $x$) range 
analyzed in this 
paper it becomes more questionable to neglect this effect. Its influence will be studied in a
future publication.

In the present study
the carriers are considered to be electrons and the kinetic term
describes a conduction band. However, in
Ga$_{1-x}$Mn$_x$As the carriers are holes 
and the valence band has to be considered instead. 
Although the latter is more complicated, the simplified treatment 
followed here should yield
similar results for both cases.\cite{bhatt,paper0,paper1}\\
\indent The local spins are assumed to be classical which
 allows the parametrization of each local spin
 in terms
of spherical coordinates: $(\theta_i,\phi_i)$. The exact-diagonalization method
for the fermionic sector is
described in this section, 
following Ref.~\onlinecite{dagotto}. The partition function can be written as:
\begin{equation}
Z=\prod_i^N \left(\int_0^\pi
d\theta_i\,\sin\theta_i\int_0^{2\pi}d\phi\,Z_g(\{\theta_i,\phi_i\})\right).
\label{eq:partition1}
\end{equation}
where $Z_g(\{\theta_i,\phi_i\})={\rm Tr}(e^{-\beta \hat{K}})$ 
is the partition function of the fermionic sector,
$\hat{K}=\hat{H}-\mu\hat{N}$, $\hat{N}$ is the number operator,
and $\mu$ is the chemical potential.\\
\indent In the following,
an hypercubic lattice of dimension $D$, length $L$, and number of sites
$N=L^D$ will be considered. Since ${\hat K}$ is a Hermitian operator, it can be represented
in terms of a hermitian matrix which can be diagonalized
by an unitary matrix $U$ such that
\begin{equation}
U^{\dagger}KU=\pmatrix{\epsilon_1&0         &\ldots&0\cr
                       0         &\epsilon_2&\ldots&0\cr
                       \vdots    &\vdots    &\ddots&\vdots\cr
                       0         &0         &\ldots&\epsilon_{2N}\cr}.
\end{equation}
The basis in which the matrix $K$ is diagonal is given by the
eigenvectors 
$u^{\dagger}_1|0\rangle$, $\ldots$, $u^{\dagger}_{2N}|0\rangle$,
where the fermionic
operators used in this basis are obtained from the
original operators through
$u_m$=$\sum_{j \sigma} U^\dagger_{m,j\sigma} 
c_{j \sigma}$, with $m$ running from 1 to $2N$.

Defining $u^{\dagger}_m u_m$=$\hat n_m$ and denoting by $n_m$ 
the eigenvalues of $\hat n_m$, 
the trace can be written
\begin{eqnarray}
 && {\rm Tr}_g(e^{-\beta {\hat K}})=\sum_{n_1,\ldots,n_{2N}} \!
 \langle n_1 \ldots n_{2N}| e^{-\beta \hat K}
 |n_1 \ldots n_{2N}\rangle \! \nonumber \\
 && = \sum_{n_1,\ldots,n_{2N}} \! \langle n_1 \ldots n_{2N}|
  e^{-\beta \sum_{\lambda=1}^{2N} \epsilon_{\lambda}n_{\lambda}}
  |n_1 \ldots n_{2N}\rangle,
\end{eqnarray}
since in the $\{u^{\dagger}_m|0\rangle \}$ basis, the operator 
$\hat K$ is $\sum_{\lambda} \epsilon_{\lambda} {\hat n}_{\lambda}$, 
and the number operator
can be replaced by its eigenvalues.
%
Thus, Eq.~(\ref{eq:partition1}) can be rewritten as
\begin{eqnarray}
  Z = \prod_i^N(\int_0^{\pi}d\theta_i \sin \theta_i \int_0^{2 \pi}d\phi_i)
  \prod_{\lambda=1}^{2N}(1+e^{-\beta \epsilon_{\lambda}}),
 \label{eq:partition}
\end{eqnarray}
\noindent which is the formula used in the simulations.

\subsection{Monte Carlo Method}
The integral over the angular variables in Eq.~(\ref{eq:partition})  
can be performed using a classical Monte Carlo simulation.\cite{montecarlo}
The eigenvalues must be obtained for each classical spin configuration
using library subroutines. Finding the eigenvalues is the most time
consuming part of the numerical simulation. 
The integrand is clearly positive. Thus, 
``sign problems'' in which the integrand of the multiple integral under
consideration can be non-positive, are fortunately not present in our study.\\
\indent Although the formalism is in the grand-canonical ensemble, 
the chemical potential was adjusted 
 to give the desired carrier density, $n$. To do so, the 
equation $n(\mu)-n=0$ was solved for $\mu$ at every 
Monte Carlo step by using the Newton-Raphson method.\cite{aliaga} 
This technique proved very efficient in adjusting with precision the
desired electronic density and, as a consequence, 
our analysis can be considered
as in the canonical ensemble as well, with a fixed number of carriers.\\ 
\indent Usually, 2,000-5,000 Monte Carlo iterations were used 
to let the system thermalize, and
then 5,000 -10,000 additional steps 
were carried out to calculate observables, 
measuring every five of these steps to reduce
autocorrelations.\\
\indent Since for fixed parameters there are many 
possibilities for the random location of Mn
impurities, results are averaged over
several of these disorder configurations.
Approximately 10-20 
disorder configurations were generally used for small lattices
 ($4^3$ and 10$\times$10) and
4-8 for larger lattices ($6^3$ and 12$\times$12). The inevitable 
uncertainties arising from the use of a small number of disorder
configuration does not seem to affect in any dramatic way our 
conclusions below. This is deduced from the analysis of results
for individual disorder samples. The qualitative trends emphasized
in the present paper  are present
in all of these configurations.

\subsection{Observables}

Quantities that depend on the Mn degrees of freedom ($\theta_i$ and $\phi_i$ in the
previous formalism) are calculated simply by averaging over the Monte Carlo configurations. 
Note that any observable that does not have the continuous symmetry of the
Hamiltonian will vanish over very long runs. 
Thus, it is standard in this context to calculate
 the absolute value of the magnetization, 
$|M|=\sqrt{\sum_{ij}\vec{S}_i\cdot \vec{S}_j}$,
as opposed to the magnetization vector. Another useful quantity is the spin-spin 
correlation, defined
by: 
\begin{equation}
C({x})=\frac{1}{N({x})}\sum_{{y}} \vec{S}_{{y}+{x}} \cdot \vec{S}_{{y}}
\end{equation}
where $N({x})$ is the number of non-zero terms in the sum. 
The correlation at a distance $d$ is averaged over
all lattice points that are separated by that distance,
 but since the system is diluted, the quantity must be normalized to the number of pairs
 of spins separated by $d$, to compare the results for different distances.\\
\indent The observables that directly depend on the electronic 
degrees of freedom can be expressed
in terms of the eigenvalues and eigenvectors of the 
Hamiltonian matrix ${\hat K}$.
 The DOS, $N(\omega)$, is simply given by
$\sum_{\lambda}\delta(\omega-\epsilon_\lambda)$. 
%
However, the majority and minority DOS, $N_\uparrow(\omega)$ and 
$N_\downarrow(\omega)$,
were also calculated in this study. 
$N_\uparrow(\omega)$ indicates the component that
aligns with the local spin, i.e., $N_\uparrow(\omega)$ is the Fourier
transform of 
$\sum_i<\tilde{c}^\dagger_{i\uparrow}(t)\tilde{c}_{i\uparrow}(0)>$, 
where 
$\tilde{c}_{i\uparrow}=\cos(\theta_i/2)c_{i\uparrow}+\sin(\theta_i/2){\rm
e}^{-i\phi_i}c_{i\downarrow}$. Then:
\begin{eqnarray} \nonumber
N_\uparrow(\omega) & = & \sum_\lambda^{2N}\delta(\omega-\epsilon_\lambda){[}\sum_i^N
U^\dagger_{i\uparrow,\lambda}U_{\lambda,i\uparrow}\cos^2(\theta_i/2)
+\\ \nonumber
& & U^\dagger_{i\downarrow,\lambda}U_{\lambda,i\downarrow}\sin^2(\theta_i/2)+\\ \nonumber
& & \left(U^\dagger_{i\uparrow,\lambda}U_{\lambda,i\downarrow}\exp(-i\phi_i)+
U^\dagger_{i\downarrow,\lambda}U_{\lambda,i\uparrow}\exp(i\phi_i)\right)\times \\ 
& & \cos(\theta_i/2)\sin(\theta_i/2){]},
\label{eq:nofomega}
\end{eqnarray}
\noindent where for sites $i$ without an impurity 
$\theta_i=\phi_i=0$ is assumed.
A similar expression is valid for $N_\downarrow(\omega)$. 
The optical conductivity was calculated as:
\begin{equation}
\sigma(\omega)=\frac{\pi(1-e^{-\beta\omega})}{\omega N}
\int^{+\infty}_{-\infty} \frac {dt}{2\pi} e^{i\omega t}
<\vec{j}_x(t)\cdot \vec{j}_x(0)>,
\end{equation}
where the current operator is:
\begin{equation}
\vec{j}_x=ite\sum_{j\sigma}(c^\dagger_{j+{\hat x},\sigma}c_{j,\sigma}-H.c.),
\end{equation}
\noindent with ${\hat x}$ the unit vector along the $x$-direction.
For $\omega\ne 0$, $\sigma(\omega)$ can be written as:
\begin{eqnarray}
\sigma(\omega) & = & \sum_{\lambda\ne\lambda'}\frac{\pi t^2 e^2 (1-e^{-\beta\omega})}{\omega
N} \times \\
\nonumber
& &
\frac{|\sum_{j\sigma}
(U^\dagger_{j+\hat{x}\sigma,\lambda}U_{j\sigma,\lambda'}
-U^\dagger_{j\sigma,\lambda}U_{j+\hat{x}\sigma,\lambda'})
|^2}{(1+e^{\beta(\rho_\lambda-\mu)})(1+e^{-\beta(\rho_{\lambda'}-\mu)})}\times\\ 
& & \delta(\omega+\rho_\lambda-\rho_{\lambda'}).
\label{eq:sigma}
\end{eqnarray}
Both $N(\omega)$ and $\sigma(\omega)$ were broadened 
using a Lorentzian function as a
substitute to the $\delta$-functions that appear in Eqs.~(\ref{eq:nofomega}) 
and (\ref{eq:sigma}).
The width of the Lorentzian used was $\epsilon=0.05$ in units 
of the hopping, $t$.\\
\indent The optical conductivity in $d$ dimensions obeys the sum rule:
\begin{equation}
\frac D2 = \frac{\pi e^2 <-\hat{T}>}{2Nd} - \int_{0^+}^\infty \sigma(\omega)\,d\omega,
\end{equation}
where $D$ is the Drude weight and ${\hat T}$ is the kinetic energy:
\begin{equation}
-\hat{T} = t\sum_{<ij>,\sigma}
(c^\dagger_{i\sigma}c_{j\sigma}+H.c.).
\end{equation}

Although the Drude weight gives a measure of the d.c. 
conductivity properties of the cluster, the 
``mesoscopic'' conductance was also calculated in order to gather additional
information. The details are explained in Section~\ref{sec:conductance}.
 
 

\section{Density of States} \label{sec:dos}

\begin{figure}
\centering{
\includegraphics[width=8cm]{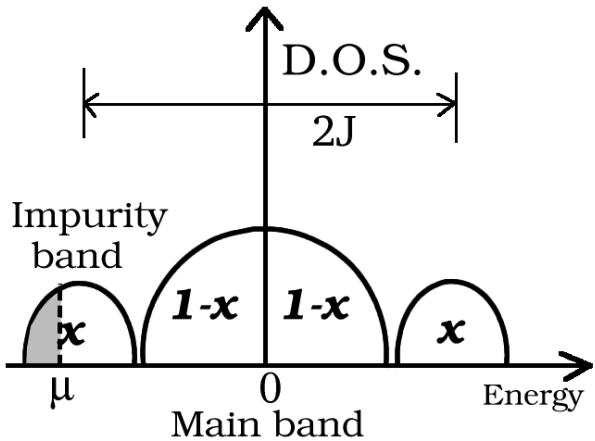}
}
\caption{Schematic representation of the DOS, for a 
ferromagnetic configuration and strong enough $J/t$ coupling. 
The ``impurity band'' has weight $x$, and the chemical 
potential $\mu$ lies within it.
\label{fig:doscartoon}}
\end{figure}

In this section, the density-of-states calculated 
at several couplings is shown and discussed. 
A few basic facts about the DOS of the model, Eq.~(\ref{eq:ham}), 
are presented first. If carrier localization effects are not taken into
account and a ferromagnetic state is considered at strong $J/t$ coupling, 
then two spin-split bands 
(``impurity bands'') will appear for this model, 
each with weight proportional to
$x$, at each side of the unperturbed band which would have weight $2(1-x)$
(Fig.~\ref{fig:doscartoon}). For partially compensated samples, $p<x$, 
the chemical potential will be located somewhere in the first spin-split band,
 and, as a consequence, only this
band would be relevant in the model Eq.~(\ref{eq:ham}).

\begin{figure}[hb]
\centering{
\includegraphics[width=8cm]{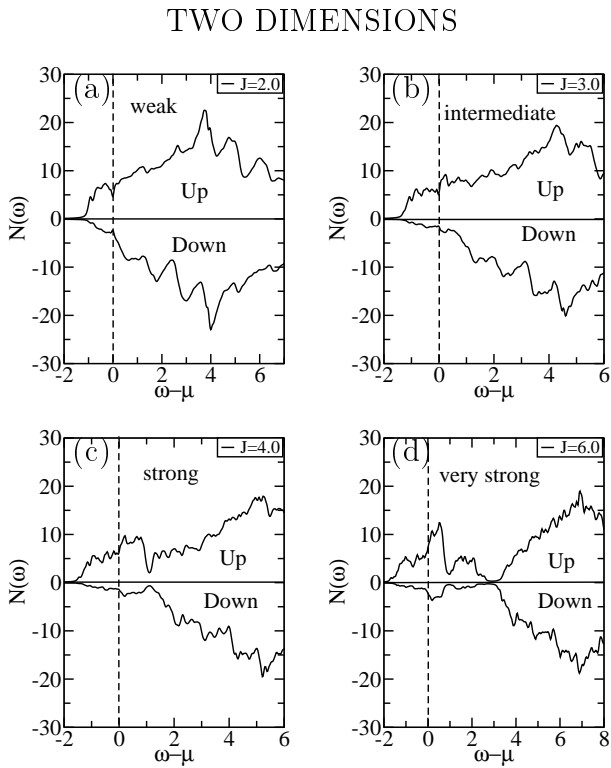}
}
\caption{$N_\uparrow(\omega)$ and $N_\downarrow(\omega)$ vs. $\omega$-$\mu$ 
for a 10$\times$10 periodic system
with 26 spins ($x\sim0.25$) and 10 electrons ($p\sim0.4$) 
at $T/t=0.01$. Results are shown for 
 (a) $J/t=2.0$, (b) $J/t=3.0$, (c) $J/t=4.0$, and (d) $J/t=6.0$. 
$\omega$-$\mu$ is in units of the hopping, $t$. 
The results are averages over eight
configurations of disorder (but there is no qualitative
difference in the results from different configurations).
\label{fig:dosparamj}}
\end{figure}

\begin{figure}[hb]
\centering{
\includegraphics[width=8cm]{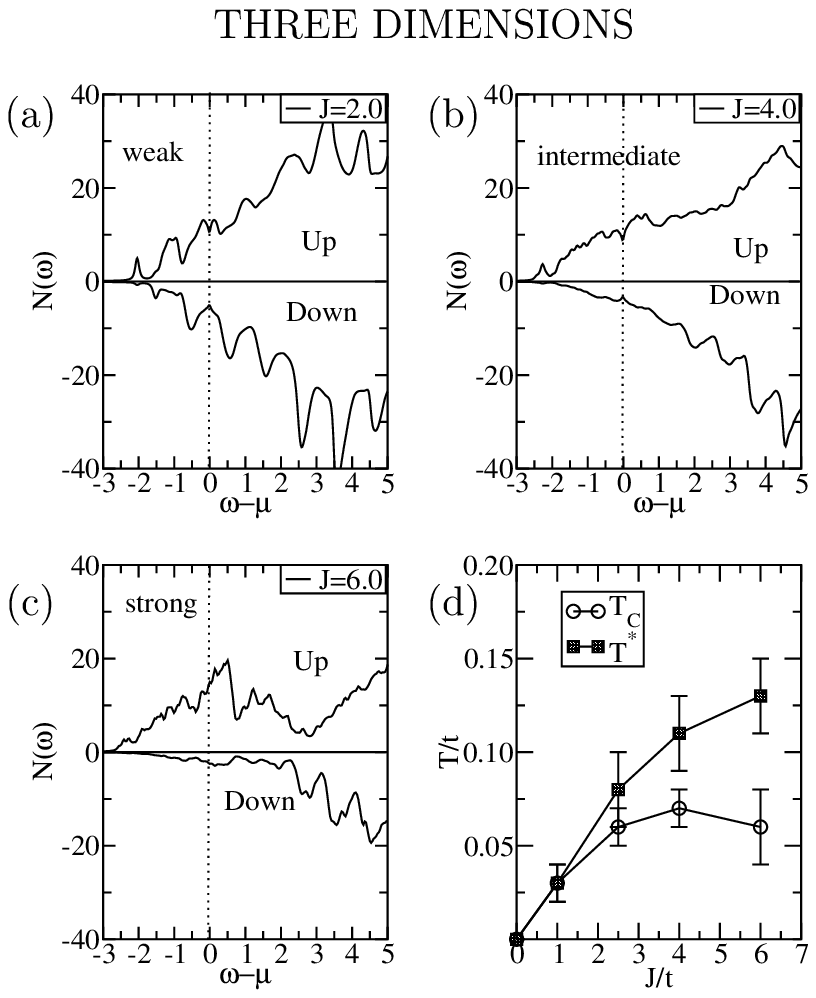}}
\caption{
$N_\uparrow(\omega)$ and $N_\downarrow(\omega)$ vs. $\omega$-$\mu$ 
for a $6^3$ periodic system
with 54 spins ($x\sim0.25$) and 16 electrons ($p\sim0.4$) at $T/t=0.01$.
Results are shown for (a) $J/t=2.0$, (b) $J/t=4.0$, (c) $J/t=6.0$. 
$\omega$-$\mu$ is in units of the hopping, $t$.
The results are averages over eight
configurations of disorder (but there is no qualitative
difference in the results from different configurations).
 (d) $T_{\rm C}$ and $T^*$ vs. $J/t$ for a $6^3$ lattice with $x=0.25$ and $p=0.4$.
 These temperatures were determined from the spin-spin correlations at short and large
 distances, as explained in Ref.~\onlinecite{paper1}. 
\label{fig:dosparamj3d}}
\end{figure}

In practice, Monte Carlo simulations show that, in the regime of
interest where $T_{\rm C}$ is optimal,
the impurity band is not completely separated from the unperturbed band.
For this reason, both the unperturbed band and the
impurity band have to be considered in quantitative calculations. 
To illustrate this, Figure \ref{fig:dosparamj} 
shows $N(\omega)$ vs. $\omega$ for
 different $J/t$'s in two dimensions.
With growing $J/t$, the ``impurity band'' begins to form for $J/t \geq 3.0$,
and at $J/t=6.0$  it is already fairly
separated from the main band. However, note that for 
this extreme regime of $J/t$, $T_{\rm C}$ is far lower than for
$J/t=3.0$, as discussed before in Ref.~\onlinecite{paper1}. 
In fact, the optimal $J/t$ for the $(x,p)$
parameters used here was found to be
in the range 2.0-4.0. This implies that the 
model shows the highest $T_{\rm C}$ 
when the ``impurity band'' is about to form, 
but it is not yet separated from the main band. This introduces 
an important difference between our approach to DMS materials, and the
theoretical calculations presented in Ref.~\onlinecite{bhatt}.
Figure \ref{fig:dosparamj3d}(a)-(c) indicates that a
similar behavior is found in three dimensions. In this case, once again, 
the optimal $T_{\rm C}$
occurs for intermediate $J/t$, as seen in Fig.~\ref{fig:dosparamj3d}(d)
where the phase diagram
obtained from a $6^3$ lattice with $x$=$0.25$ and $p$=$0.3$ is shown.\\
\indent The physical reason for this behavior is that at very 
large $J/t$, the states
of the impurity band are highly localized where the Mn spins are. Local ferromagnetism can
be easily formed, but global ferromagnetism is suppressed by the concomitant
weak coupling between magnetized clusters.

Both, Figures \ref{fig:dosparamj} and  \ref{fig:dosparamj3d} represent results
obtained at particular values of parameters  
$x$=$0.25$, $p$=$0.3$, and $T/t=0.01$. However, results for several 
other sets $\{ (x,p,T) \}$
were gathered as well (not shown).
The $x$ dependence of the DOS is simple: increasing $x$ 
produces a proportional increase in the number of states 
of the impurity band, and a corresponding decrease of the
main band weight (see Fig.\ref{fig:doscartoon}).
In terms of $p$, increasing the effective carrier 
concentration by a small amount 
was found to simply shift the chemical potential to the right. Concerning the temperature
dependence of the DOS, at low temperature the states contributing to the impurity band are
polarized, i.e. the system is ferromagnetic, 
as shown by the different weights of the majority and minority bands 
(Figs.~\ref{fig:dosparamj} and \ref{fig:dosparamj3d}).
 As the temperature increases, spin disorder grows due
to thermal fluctuations, and when $T\sim T^*$, i.e. in the 
 paramagnetic regime, the bands become symmetric.


It appears that the only change that
a small increase in the carrier concentration produces is an increase in the
chemical potential. However, experimentally 
the appearance of a pseudo-gap at the Fermi energy has been reported.\cite{chun}
Although the finite size effects in the present theoretical calculation
do not allow for a detailed study of $N(\omega)$ at or very near the Fermi energy 
with high enough precision, in certain cases a pseudo-gap was indeed observed in the present
study at the chemical potential, particularly when the system is in the clustered regime.
While this result certainly needs confirmation, it is tempting to draw analogies 
with the more detailed calculations carried out in the context of manganites,
where the presence of a pseudo-gap both in theoretical models and in experiments
is well established.\cite{moreo,dagotto,book} Pseudo-gaps are also present in under-doped
high temperature superconductors. Given the analogies between DMS materials
and transition-metal oxides unveiled in previous investigations,\cite{paper1}
it would not be surprising that DMS presents a pseudo-gap in the clustered
regime as well. More work is needed to confirm these speculations.

Results presented in this section for the DOS 
qualitatively agree with those found in Ref.~\onlinecite{hwang}
 using the dynamical mean-field technique (DMFT) 
in the coupling regime studied there. However, the DMFT approach 
is a mean-field approximation local in space and, as a consequence, the state
 emphasized in this paper --with randomly distributed clusters-- cannot be studied
 accurately with such a technique.

Experimental evidence of the formation of the impurity band for Ga$_{1-x}$Mn$_x$As
has been provided by Okabayashi {\it et al.} 
in Refs.~\onlinecite{okabayashi1}-\onlinecite{okabayashi2} 
(see also Ref.~\onlinecite{photo}). It is
interesting to remark that half the total width of the impurity band obtained in our simulations is
about $2t$ (see, e.g., Figs. \ref{fig:dosparamj} and \ref{fig:dosparamj3d}), and using
 $t$=$0.3$eV\cite{paper1} this is equal
to 0.6eV, in good agreement with the width 0.5eV
 estimated from the photoemission experimental measurements
mentioned before.\cite{okabayashi1}
\begin{figure}
\centering{
\includegraphics[width=5cm]{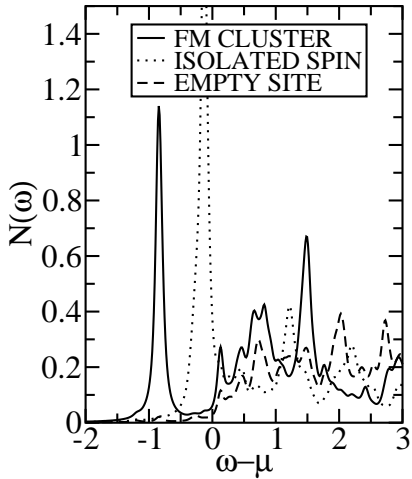}}
\caption{Local DOS for three different classes of sites (see text) on an 
8$\times$8 lattice with $J/t=2.5$, $T/t=0.01$, $x\sim0.2$ and $p=0.4$.
This coupling regime corresponds to the intermediate optimal region, where the
impurity band is not fully formed. ``FM CLUSTER'' is $N_{\rm I}(\omega)$ 
in the text notation, while ``ISOLATED SPIN'' and ``EMPTY SITE'' are
$N_{\rm II}(\omega)$ and $N_{\rm III}(\omega)$, respectively.
\label{fig:ldos}}
\end{figure}

The local DOS is studied next. The purpose of this analysis is to further 
understand the inhomogeneous state that forms as a consequence 
of Mn-spin dilution and concomitant carrier localization. 
The local DOS is shown in Fig. \ref{fig:ldos} on an 8$\times$8 lattice at
$x=0.25$, $p=0.4$, $J/t=2.5$, and $T/t=0.05$. 
In this particular case only, an 
arrangement of classical spins was introduced `by hand'
such that the clusters are clearly formed. 
In this way, sites can easily be classified in three groups, 
as discussed below. Despite using a
particularly chosen spin configuration, the resulting DOS 
is similar to those obtained previously (Fig.\ref{fig:dosparamj}) using
a truly random distribution of spins. Lattice sites are 
classified as follows: the first group is defined 
to contain lattice sites that have a classical spin and 
belong to a spin cluster. The second class is
composed of sites that have an isolated  
classical spin, i.e. nearby sites do not have other
classical spins. Finally, lattice sites without classical spins in the
same site or its vicinity belong to the third class. 
$N_{\rm I}(\omega)$, $N_{\rm II}(\omega)$, and $N_{\rm III}(\omega)$  
denote the local DOS at sites corresponding to each of 
the three classes, respectively. It can be observed that  
$N_{\rm I}(\omega)$ contributes 
to states inside the impurity band (note the sharp peak near -1), but also
contributes to the main band since the $J/t$ used is intermediate.
$N_{\rm II}(\omega)$ contributes 
a sharp peak near zero, and also has weight in the main band. 
This can be explained
as follows:  the electron is weakly localized  at an isolated 
site yielding a state at $\omega_{\rm II}\approx -J$ or, 
since $\mu\approx -J$ for $p\approx0.4$, $\omega_{\rm II}-\mu\approx 0$ 
(see Fig.\ref{fig:ldos}a). This interesting result shows that many 
of the states near zero in Figs.\ref{fig:dosparamj} and \ref{fig:dosparamj3d}
may be {\it localized}, and they do not contribute to the conductivity.
Finally, in 
$N_{\rm III}(\omega)$ the empty sites contribute weight only in nearly 
unperturbed states, i.e., states that belong to the main band. In conclusion,
 the local carrier density is very inhomogeneous  and 
 this fact has an
 important effect on the form of the site-integrated DOS. Our results 
show that scanning tunneling microscopy (STM) experiments would be able 
to reveal the clustered structure proposed here, if it indeed exists,
when applied to DMS materials.

\section{Optical Conductivity} \label{sec:sigma}

\begin{figure}[h]
\includegraphics[width=9cm]{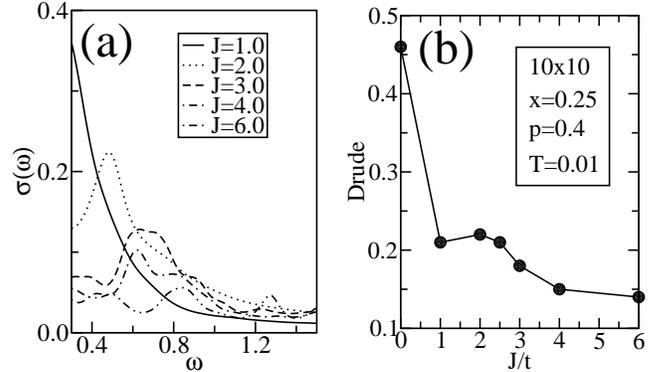}
\caption{Coupling dependence of the conductivity 
at low temperature in two dimensions. 
(a) $\sigma(\omega)$ vs. $\omega$ 
for a 10$\times$10 periodic system
with 26 spins ($x\sim0.25$), 10 electrons ($p\sim0.4$), 
$T/t=0.01$, and for different $J/t$'s as shown. 
(b) Drude weight, $D$, vs. $J/t$ for the same lattice and parameters as in (a).
\label{fig:swparamj}}
\end{figure}

\begin{figure}[h]
\includegraphics[width=9cm]{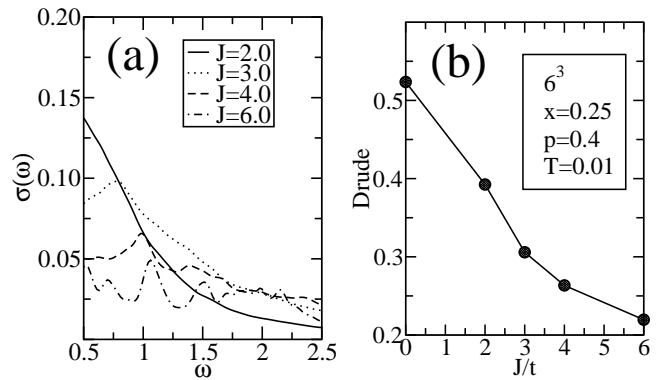}
\caption{Coupling dependence of the conductivity 
at low temperature in three dimensions.
(a) $\sigma(\omega)$ vs. $\omega$ for a $6^3$ periodic system
with 54 spins ($x\sim0.25$), 16 electrons ($p\sim0.4$), $T/t=0.01$,
and for different $J/t$'s as shown.
 (b) $D$ vs. $J/t$ for the same lattice and parameters as in (a). 
\label{fig:swparamj3d}}
\end{figure}

In the previous section, it was argued that at the
optimal coupling $J/t$ -- where $T_{\rm C}$ is maximized -- 
the impurity band is not completely 
separated from the main band. This conclusion is supported by
results obtained from the optical conductivity as well, 
which is shown in Fig.~\ref{fig:swparamj}(a) 
for a two-dimensional lattice. This optical conductivity has two main
features: (i) a zero-frequency or Drude peak and (ii) a 
finite-frequency broad peak which is believed to correspond
to transitions from the 
impurity band to the main band, as argued below. From Fig.~\ref{fig:swparamj}(a),
it is observed that inter-band transitions are not much relevant
 at weak coupling ($J/t$=$1.0$), but they appear with more weight at
intermediate couplings ($J/t$=$2.0-3.0$), where $T_{\rm C}$ is optimal. 
It is worth remarking that at $J/t$ strong enough, e.g.
$J/t$=$6.0$, when the impurity band is well-formed, 
the finite-frequency peak is weaker in strength than for
the optimal $J/t$ due to localization. 
In fact, if $J/t$ were so large that
the ``impurity band'' is completely separated 
from the unperturbed band, then carrier localization 
would be so strong that inter-band transitions would not be possible.
The Drude peak is
plotted separately in Fig.~\ref{fig:swparamj}(b) 
as a function of $J/t$ at low temperatures. 
As expected, when $J/t$ increases and localization sets in, 
the conductivity of the cluster decreases. 
Similar results for the optical conductivity and
Drude weight are found in three dimensions (see Fig.~\ref{fig:swparamj3d}).

The rise in absorption that appears for $J/t\ge2.5$ at intermediate 
frequencies is due to inter-band transitions, i.e. transitions
from the impurity band to the unperturbed band. 
The frequency of this peak, $\omega_{inter}$, depends upon $J/t$ 
as well as carrier concentration $p$, but for 
$J/t\ge2.5$ (which is an intermediate value) and $p=0.4$ (a realistic value),
it was observed that $0.5t\le\omega_{inter}\le1.0t$. The
model parameter $t$ was
previously estimated\cite{paper1} to be $t$$\sim$$0.3eV$, yielding
$0.15eV\le\omega_{inter}\le0.3eV$, in agreement with experiments (see below).

\begin{figure}[h]
\includegraphics[width=7.5cm]{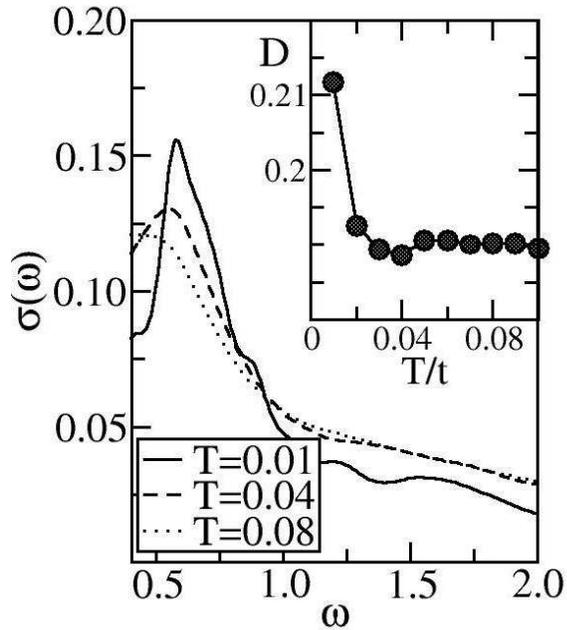}
\caption{Temperature dependence of the optical conductivity 
of model Eq.~(\ref{eq:ham}). 
Shown is $\sigma(\omega)$ vs. $\omega$ for a 
10$\times$10 periodic system
with 26 spins ($x\sim0.25$), $J/t=2.5$, $p=0.3$, and  
for different temperatures, as indicated. 
{\it Inset}: Drude weight, $D$, vs. temperature $T$.\label{fig:swt}}
\end{figure}

 The temperature dependence of $\sigma(\omega)$ is shown in 
Fig. \ref{fig:swt}, 
and it is as follows: for $T\le T_{\rm C}$, the Drude weight, $D$, 
decreases in intensity as temperature increases and the same behavior is
observed for the finite-frequency peak. 
 When $T\ge T_{\rm C}$, $D$ increases slightly at first and then
remains constant as the temperature is increased further. 
This slight raise in $D$ is subtle
and could be linked to the decrease in resistivity 
observed for $T>T_{\rm C}$ for intermediate 
$J/t$ as explained in the next section, since the system is
clustered just above $T_{\rm C}$. Since both the optical
conductivity and Drude weight decrease with temperature, 
to satisfy the sum rule the
kinetic energy must also decrease as temperature increases. 
This is indeed observed (not shown) and is to be expected since 
localization of the wave function is stronger above $T_{\rm C}$.

\begin{figure}[h]
\includegraphics[width=7cm]{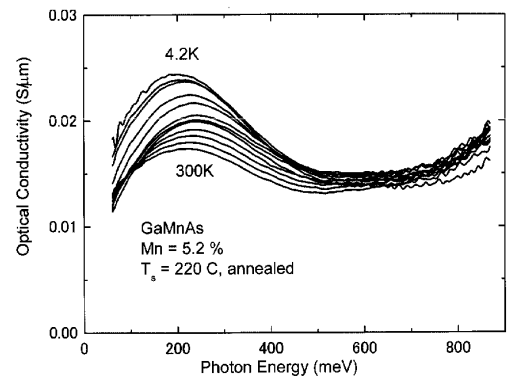}
\caption{Absorption coefficient $\alpha(\omega)$ spectra for a metallic sample
 prepared by low temperature annealing. The temperatures from down to top are
300, 250, 200, 150, 120, 100, 80, 60, 40, 20, 10 and 4.2 K. 
(from Katsumoto {\it et al.}\cite{katsumoto})\label{fig:katsumoto}
}
\end{figure}

A broad peak at around $0.2$eV is experimentally observed in the
optical conductivity of Ga$_{1-x}$Mn$_x$As, 
as shown in Fig.~\ref{fig:katsumoto} which is reproduced 
from Ref.~\onlinecite{katsumoto}. This feature  has been explained 
before in two different ways: (i) as produced by transitions
from the impurity band to the GaAs valence band\cite{hwang} or (ii) as caused by
inter-valence band transitions.\cite{sinova} Due to the frequency 
range and temperature behavior observed,
our study supports the first possibility, 
i.e., that the finite-frequency peak observed
for Ga$_{1-x}$Mn$_x$As at around $0.2$ eV is due to transitions 
from the impurity band to the main band.
Recent observations\cite{basov} also appear to support this view.

\begin{figure}[h]
\includegraphics[width=7.5cm]{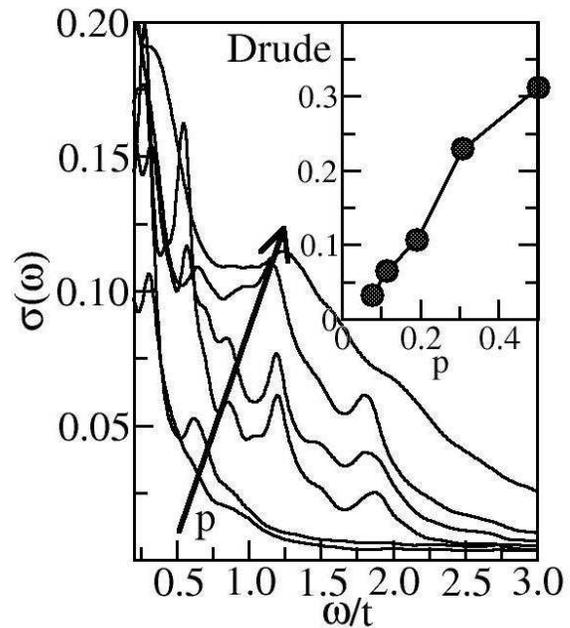}
\caption{$p$-dependence of the optical conductivity. 
Shown is $\sigma(\omega)$, including the Drude weight, 
vs. $\omega$ for a 10$\times$10 lattice, $x=0.25$, $J/t=2.5$, $T/t=0.01$,  
and different values of $p$.
In the direction of the arrow, $p$ takes the values 0.08, 0.12, 
0.20, 0.30, 0.38, and 0.5. {\it Inset}: Drude weight, $D$, vs. 
fraction of carriers, $p$. 
\label{fig:swp}}
\end{figure}

The Drude weight increases in intensity with increasing
carrier density, as can be seen from Fig.~\ref{fig:swp}.
In fact, the ratio $p$ of carriers to Mn 
concentration, which is a measure of the compensation of the samples, could
explain, within the framework of this model, the different behavior 
observed at low frequencies 
for GaMnAs and InMnAs. In the former case, no tail of the Drude
peak is found.
However, for InMnAs, a clear Drude tailing 
is observed,\cite{hirakawa,katsumoto}
with increasing intensity as the temperature increases.  At
large enough doping, our calculations based on model Eq.~(\ref{eq:ham}) predict
a Drude-like peak for $\sigma(\omega)$ which 
could correspond to the regime valid for InMnAs, 
while for low doping this study predicts a 
very small Drude peak, which is the case for GaMnAs. 
Unfortunately, the precise carrier concentrations
for both materials are not known experimentally with precision.

\section{Conductance and Metal-Insulator Transition} \label{sec:conductance}

\begin{figure}
\includegraphics[width=8cm]{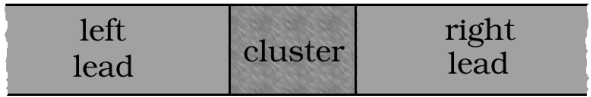}
\caption{Geometry used for the calculation of the conductance. 
The interacting region (cluster) is connected by
ideal contacts to semi-infinite ideal leads.\label{fig:leads}}
\end{figure}

The conductance, $G$, is calculated  
using the Kubo formula adapted to geometries usually employed in the
context of mesoscopic systems.\cite{verges} The actual expression is
\begin{equation} 
G=2\frac{e^2}{h} \mathrm{Tr}\left[(i\hbar \hat{v}_x) \mathrm{Im} \hat{\mathcal{G}}(E)
 (i\hbar \hat{v}_x) \mathrm{Im} \hat{\mathcal{G}}(E)\right],
\label{eq:conductance}
\end{equation} 
where $\hat{v}_x$ is the velocity operator in the $x$ direction and 
$ \mathrm{Im} \hat{\mathcal{G}}(E)$ is obtained from the advanced and retarded
 Green functions using
$2i \mathrm{Im} \hat{\mathcal{G}}(E)=\hat{\mathcal{G}}^R(E)-\hat{\mathcal{G}}^A(E)$, where $E$
is the Fermi energy. The cluster is considered 
to be connected by ideal contacts to 
two semi-infinite ideal leads, as represented in Fig.~\ref{fig:leads}.
Current is induced by an infinitesimal voltage drop.
 This formalism avoids some of the  
problems associated with finite systems, such as
 the fact that the conductivity is given by a 
sum of Dirac $\delta$ functions, since the matrix eigenvalues are
discrete. A zero-frequency delta peak with finite weight corresponds
 to an ideal 
metal -- having zero resistance -- unless an arbitrary width is given
to that delta peak. In addition, in numerical studies sometimes
it occurs that the weights
of the zero-frequency delta peak are negative due to size 
effects.\cite{RMP} 
For these reasons, calculations of d.c. resistivity using finite
close systems are rare in the literature. All these
problems are avoided with the formulation described here.

\begin{figure}[h]
\includegraphics[width=7cm]{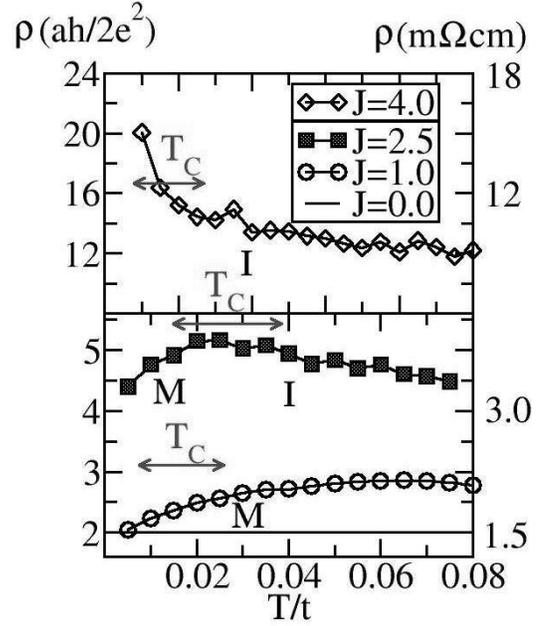}
\caption{Dependence of the theoretically calculated resistivity, $\rho$,
with temperature in three dimensions.
Shown is  $\rho=L/G$ vs. $T$ on $4^3$ lattices, 16 spins ($x=0.25$), and 5 carriers
($p=0.3$) for the $J/t$'s indicated. An average over 20 disorder configurations has been
performed in each case. Units are shown in two scales, $a h/(2e^2)$ on the left and 
m$\Omega$cm on the right, with $L=4$ and assuming \mbox{$a=5.6$ \AA}.
 The estimated critical temperatures are also shown (the arrows indicate the current accuracy of the estimations).
\label{fig:rho4d3}}
\end{figure}

The entire equilibrated cluster as obtained from the MC simulation is introduced in the
geometry of Fig.~\ref{fig:leads}. The ideal leads enter the formalism through
self-energies at the left/right boundaries, as described in Ref.~\onlinecite{verges}.
In some cases a variant of the method explained 
in Ref.~\onlinecite{verges} was 
used in this paper to calculate the conductance (this modification
was suggested to us by J. Verg\'es). 
Instead of connecting the 
cluster to an ideal lead with equal hoppings, a replica of the cluster 
was used at the sides. This method, although slightly slower, takes into 
account all of the Monte Carlo data for the cluster, 
including the periodic boundary conditions.
In addition, averages over the random Mn spin
distributions are carried out. The
physical units of the conductance $G$ 
in the numerical simulations are $e^2/h$ as can be
inferred from Eq.~(\ref{eq:conductance}). In 
Figure~\ref{fig:rho4d3} 
the inverse of the conductance, which is a measure of the resistivity, 
is plotted for a
three-dimensional lattice at weak, intermediate, and strong coupling,
at fixed $x$=$0.25$ and $p$=$0.3$. For the weak 
coupling regime ($J/t$=$1.0$) the system is weakly metallic 
at all temperatures. In the other limit of strong couplings,
$1/G$ decreases with increasing temperature, indicating 
a clear insulating phase, as a
result of the system being in a clustered state at the temperatures 
explored,\cite{paper1,paper0} with carriers localized near the Mn spins. 
At the important intermediate couplings emphasized in our effort, 
the system behaves like a dirty metal for $T<T_{\rm C}$, while for
$T_{\rm C}<T<T^*$, $1/G$ slightly decreases with increasing temperature, 
indicating that a soft metal to
insulator transition takes place near $T_{\rm C}$.  
For $T>T^*$, where the system is paramagnetic, $1/G$ is almost constant. 
Note that for strong enough $J/t$, $T_{\rm C}\rightarrow0$ 
and therefore no metallic phase is present. 

\begin{figure}[h]
\includegraphics[width=6.0cm]{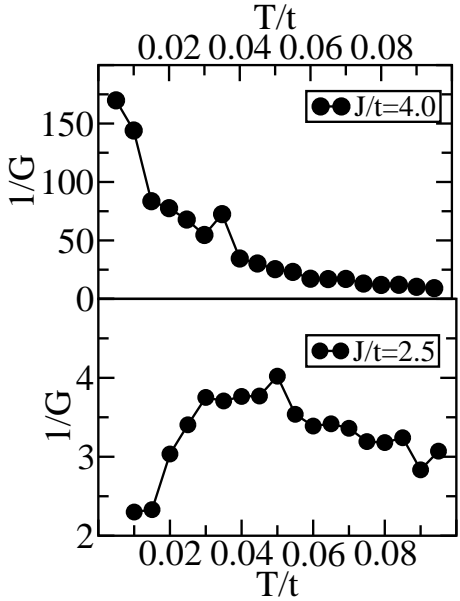}
\caption{Inverse of the conductance, $1/G$, vs. $T$ calculated 
on a 10$\times$10 lattice with 26 spins ($x\sim0.25$), 
8 electrons ($p\approx0.3$), and two values of $J/t$ as indicated. Shown is an 
average over three disorder configurations. $1/G$ has
units of $h/(2e^2)$ in two dimensions.\label{fig:rho10x10}}
\end{figure}

\begin{figure}
\includegraphics[width=6.5cm]{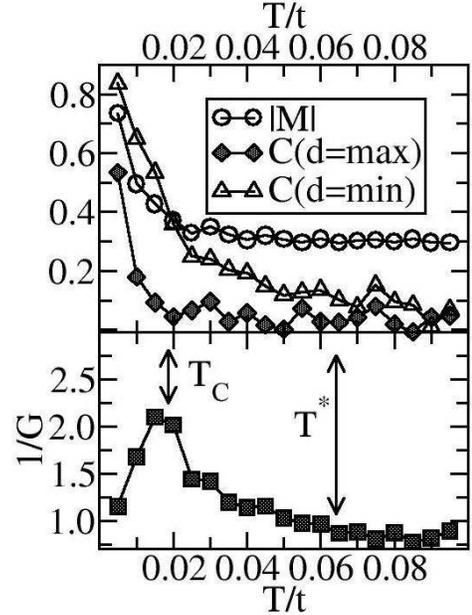}
\caption{Magnetization, $|M|$, spin-spin correlations, $C(d)$, 
at maximum and minimum distance,  
and inverse of the conductance, $1/G$, vs. $T$ 
on a 12$\times$12 lattice with 22 spins ($x\approx0.15$), 
6 electrons ($p\approx0.3$), and $J/t$=$1.0$. The approximate vanishing 
of spin-correlations at different temperatures, depending
on the distance $d$ studied, allowed us to obtain an
approximate determination of $T_{\rm C}$ and $T^*$, similarly as carried out
in previous studies.\cite{paper1}
\label{fig:rho12x12}}
\end{figure}

Similar qualitative behavior is found for the two-dimensional case
(Figs.~\ref{fig:rho10x10} and \ref{fig:rho12x12}).
Furthermore, in Fig.~\ref{fig:rho12x12} the
spin-spin correlations have been plotted to show the location 
of $T_{\rm C}$ and $T^*$ and their relation to the resistivity.


It is interesting to compare our results with experiments. 
For Ga$_{1-x}$Mn$_x$As, data similar to those found in our
investigations have been reported.\cite{katsumoto,ohno}
A typical result for 
resistivity versus $T$ can be found in Ref.~\onlinecite{potashnik}.
The qualitative behavior of the resistivity in these 
samples agrees well with the theoretical results presented 
in Fig.~\ref{fig:rho4d3} if intermediate couplings are considered. 
On the other hand, stronger or weaker couplings are not useful
to reproduce the data, since the result is either insulating or metallic
at all temperatures, respectively (defining 
metallic and insulator regimes by
the slope of the resistivity vs. temperature curves).


Furthermore, even the experimental numerical values 
of the resistivity are in agreement with the results
presented here (even though model Eq.~(\ref{eq:ham}) does not 
include a realistic treatment of the GaAs host bands).  This
can be shown as follows.
The conductivity of a three-dimensional sample is 
related to the conductance by
$\sigma=G/L$, where $L$ is the side length 
of the lattice. Hence the resistivity 
is $\rho=L/G$. The units of $G$ are, as explained before, $h/(2e^2)$, 
and in our case $L=4a$, 
where $a$ is the lattice spacing. Assuming $a=5.6$ \AA, 
then the values shown in Fig.~\ref{fig:rho4d3} are
obtained. 
%
Note that in Fig.~\ref{fig:rho4d3} 
the minimum resistivity for $J/t$=$2.5$ is 3.3~m$\Omega$cm, 
whereas the minimum possible value for that carrier doping and
cluster size used is
1.5~m$\Omega$cm, which corresponds to the case $J/t$=$0$. In spite
of the label `metallic' for these results, the absolute values of
the resistivity are high.  
Similarly, in the metallic
phase of the sample shown in the experimental 
results of Ref.~\onlinecite{potashnik}, 
the minimum resistivity is only $\sim3-6$~m$\Omega$cm. Both in theory
and experiments, the metallic phase appears to be `dirty', which is
likely due to the reduced number of carriers, and localizing effect
of the disorder.

\begin{figure}[h]
\includegraphics[width=7.5cm]{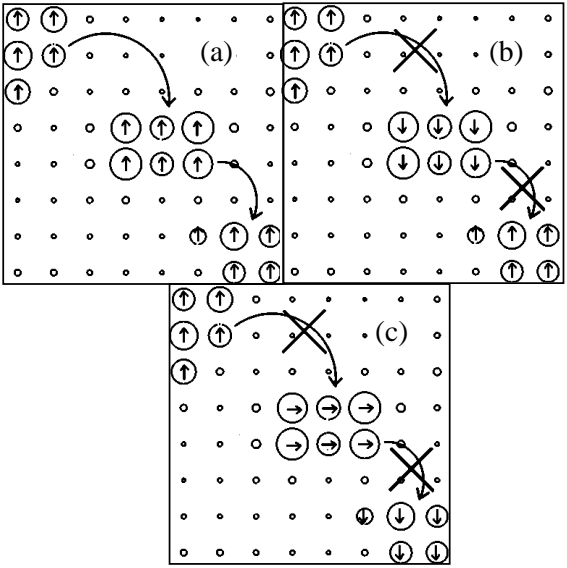}
\caption{Qualitative explanation of the transport properties 
of clustered states. Shown are three clusters created `by hand' 
on an 8$\times$8 lattice and with spin  configurations
also selected `by hand' to illustrate our ideas.
For a (a) FM state at $J/t$=$2.5$ and $p$=$0.4$, conduction is possible due to the
alignment of magnetic moments and the conductance was found to be
$G$=$2.8 (2e^2/h)$. For the ``clustered'' state regime, two 
typical configurations (b) and (c) are shown where the conduction
channels are broken and as a consequence $G\approx0$. 
For the same clustered state but with randomly selected 
spin orientations (not shown)  the conductance is small but finite
$G\approx0.5 (2e^2/h)$. The radius of the solid circles 
represent the local charge density.
\label{fig:conduction}}
\end{figure}

For $J/t$ intermediate or strong, localization of the wave function is
observed at intermediate temperatures.\cite{paper1} 
This implies that carriers tunnel between impurity sites 
without visiting the main band, leading to insulating behavior.
Below $T_{\rm C}$ conduction is favored by the 
ferromagnetic order and, as a consequence, as temperature decreases
conductance increases. 
To further understand why the conductance (resistance) has a minimum (maximum)
 around $T_{\rm C}$ 
it is helpful to consider the three states: FM, clustered, and
paramagnetic (PM) as depicted for a special spin configuration in 
Fig.~\ref{fig:conduction} on an 
8$\times$8 lattice. This configuration is not truly 
random but it was chosen `by hand' for simplicity so that the spin
clusters can easily be recognized. In the case of a 
random configuration of spins a
similar reasoning can be drawn. For the FM state 
of Fig.~\ref{fig:conduction}, conduction is possible and
indeed the measured conductance is $G\approx2.8 (2e^2/h)$. 
For the two possible clustered states
represented in Figs.~\ref{fig:conduction}(b) and (c), the conduction is reduced
due to the different alignment of spins in different clusters. 
This is verified by calculating the conductance which is $G\approx0.0$ 
in those cases. For a 
paramagnetic spin arrangement with spin pointing in random directions
(not shown), the conductance is small but finite, $G\approx0.5 (2e^2/h)$. 
It follows that the clustered state has the
minimum conductance, and this explains 
the observed behavior of the resistivity
 with a maximum near $T_{\rm C}$, i.e., when the system is clustered.

\begin{figure}[h]
\includegraphics[width=5.5cm]{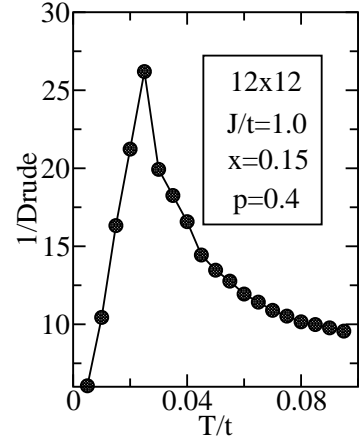}
\caption{Inverse of the Drude weight  vs. $T$ for a 12$\times$12 lattice
and parameters as in Fig.~\ref{fig:rho12x12}. Note that both the 
inverse conductance in Fig.~\ref{fig:rho12x12},  
and $Drude^{-1}$ provide evidence
for a metal-insulator transition at approximately 
the same temperature $T_{\rm C}$.\label{fig:drude}}
\end{figure}

The inverse of the Drude weight, $D^{-1}$, also shows a peak 
around $T_{\rm C}$ as seen in Fig.~\ref{fig:drude}, where the same 
Mn-spin configuration as in Fig.~\ref{fig:rho12x12} was used.
This provides further evidence for 
the metal-insulator transition near $T_{\rm C}$ 
described here. 

\begin{figure}
\centering{
\includegraphics[width=7cm]{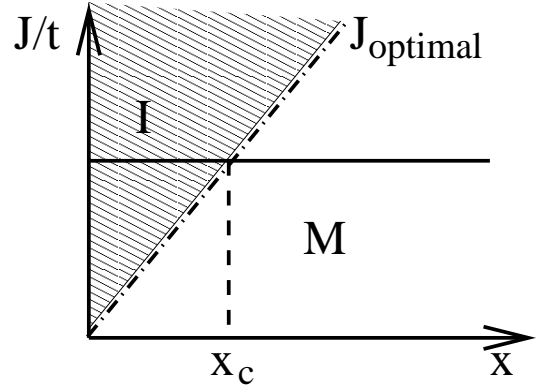}
}
\caption{Dependence of $J/t_{\rm optimal}$ with $x$, approximately
separating the metal from the insulator (standard notation). 
At fixed $J/t$ (horizontal
solid line), as in experiments, the system should transform
from an insulator to a metal with increasing $x$, in agreement with
experiments\cite{potashnik}.
\label{fig:jvsx}}
\end{figure}

It is also interesting to remark that the optimal value of $J/t$
changes with $x$ (e.g., for small $x$, $J/t$ optimal is also small).
Since $J/t_{\rm optimal}$ approximately separates the metal from the
insulator, working at a fixed $J/t$ (as in real materials) and varying
$x$, then an insulator at high temperatures is found at small $x$,
turning into a metal at larger $x$. This is in excellent agreement
with the experimental results of Ref.~\onlinecite{potashnik} using a careful annealing
procedure (while results of previous investigations with as grown samples had found an
insulator-metal-insulator transition with increasing $x$.\cite{ohno}).
The issue discussed in this
paragraph is visually illustrated in Fig.~\ref{fig:jvsx}, for the benefit of the reader.

\section{Magnetoresistance} \label{sec:magnet}

\begin{figure}[h]
\includegraphics[width=6cm]{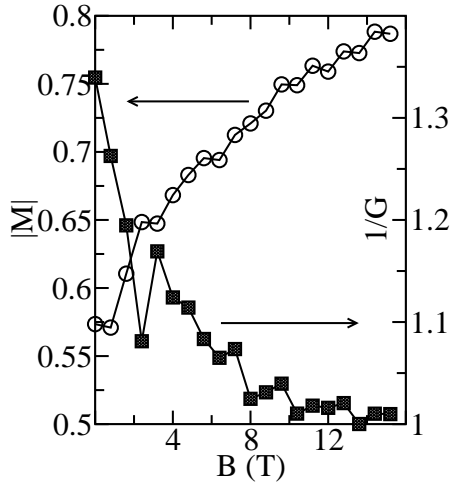}
\caption{Magnetization, $|M|$, and inverse conductance,  
$1/G$, vs. magnetic field, 
$B$, on a 12$\times$12 
lattice with PBC, $x$=$0.15$, $p$=$0.3$, $J$=$1.0$, and $T/t$=$0.01$
(same parameters as in Fig.~\ref{fig:rho12x12}). 
The units of $B$ are Tesla, assuming
$g$=$2.0$ and $S$=$5/2$. The units of $1/G$ are $h/2e^2$. 
The approximate values of the 
magnetoresistance are 18\%
at 4 Tesla,  and 30\% at 12 Tesla.\label{fig:l12x12field}}
\end{figure}

The magnetoresistance percentage ratio was calculated using the definition
$MR=100\times(\rho(0)-\rho(B))/\rho(B)$. Figure~\ref{fig:l12x12field} 
shows the magnetization, $|M|$, and the
resistivity as a function of the applied field $B$ on a 12$\times$12 lattice, 
$x$=$0.15$, $p$=$0.4$, $J/t$=$1.0$, and low temperature. 
Note that in two dimensions at $x$=$0.25$ and $p$=$0.3$, 
the intermediate coupling corresponds to $J/t$=2.5, but at
$x$=$0.15$, now $J/t$=$1.0$ corresponds to optimal coupling since as $x$ decreases 
the optimal $J/t$ also decreases.
For these parameters, $T_{\rm C}$ was estimated to be 
0.02$t$ and $T^*\approx$0.06$t$. The value in Teslas of the
magnetic field was calculated assuming $g$=$2.0$ and $S$=$5/2$ for the
localized spins. The units of the resistivity are $h/(2e^2)$ 
for the two-dimensional lattice.

In our simulations it was observed that $|M|$ increases with increasing 
magnetic field. At zero magnetic field $|M|$ is 60\% of its maximum value,
while at 12 T it has reached $\sim$80\%. This is in agreement with the 
experimental results in Fig.~3 of Ref.~\onlinecite{ohno} where
the magnetization is 50\% of its maximum value at zero field,
but it reaches 70\% of that maximum value at 4 T.
In our studies it was also observed that $1/G$ decreases  with increasing
magnetic field, showing at all fields a negative magnetoresistance.
Near $T_{\rm C}$ the resistivity decreases by 20\%-30\% increasing the field 
from zero to 12 T, while the decrease in resistivity at higher
temperatures is much smaller. 
The present computational results agree very well with the experimentally 
measured dependence of the 
resistivity and magnetization on magnetic 
fields. For example, in Fig.~2(b) of Ref.~\onlinecite{ohno}, 
the decrease in resistivity is 25\%
near $T_{\rm C}\approx60$ K when increasing the field from zero to 15 T. A  
much smaller decrease in resistivity is observed at
higher temperatures ($T>120$ K). It is also indicated in that 
experimental reference that the
magnetoresistance is between 30\% and 40\% at 12 Tesla.

It is interesting to remark that
substantial magnetoresistance effects have been reported in
thin films consisting of nanoscale Mn$_{11}$Ge$_{8}$ ferromagnetic
clusters embedded in a dilute ferromagnetic semiconductor matrix.\cite{park}
The characteristics of these materials are analogous to our clustered state,
that also show a robust MR effect due to magnetization 
rotation of spontaneously formed clusters.

\section{More about Cluster Formation} \label{sec:cluster}

\begin{figure}[h]
\centering{
\includegraphics[width=7.5cm]{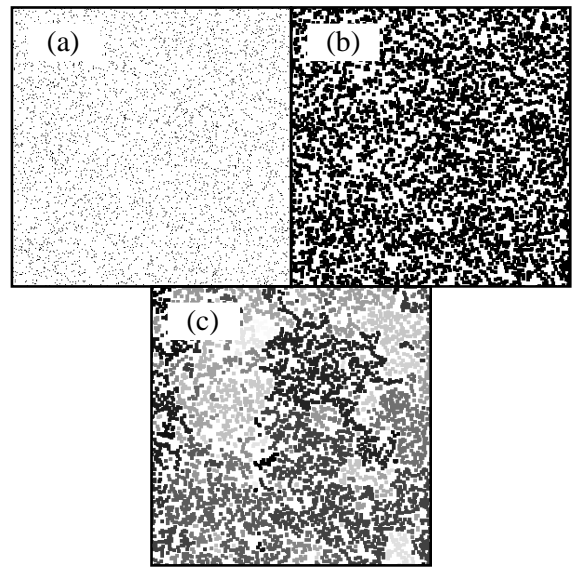}
}
\caption{Percolation picture for the formation of the clustered-state regime.
(a) Two-dimensional lattice representing randomly-located classical spins 
as black dots with $x$=$0.05$. (b) Black areas represent non-zero carrier wave function,
assuming a step-function profile for the wave functions 
with radius equal to 2 sites, as explained in the text.
(c) Same as (b) but now showing connected regions (which could in practice 
correspond to the FM clusters discussed in the text)
 indicated with different shades of gray.\label{fig:percolation}}
\end{figure}

It was discussed before in this paper and in previous publications 
that a clustered state is formed above $T_{\rm C}$
for intermediate and large $J/t$ couplings. This state is a candidate
to describe DMS materials since it explains both the resistivity maximum around
$T_{\rm C}$, as well as the decrease in resistivity with increasing applied 
magnetic field. In addition, it provides an optimal $T_{\rm C}$.
This clustered state is formed only for intermediate 
or larger $J/t$ and when the
compensation is strong, $p<0.4$. Since for large $J/t$ 
the carriers are localized, 
the problem becomes one of percolation theory, which has already been treated
using different approaches and approximations.\cite{kaminski} Here, only 
a very simple way of visualizing this state
will be presented. First, consider a two-dimensional lattice with 5\% Mn spins
represented as black dots, as shown in Fig.~\ref{fig:percolation}a. 
The carrier wave function, $\psi(r)$, is considered 
to be localized around a Mn 
spin and it is assumed to be a step function for simplicity, i.e.
$\psi(r)$ is non-zero only if $r<r_l$ where $r_l$ is the localization radius
introduced by hand. Sites where the wave function is not zero are 
the black areas of Fig.~\ref{fig:percolation}(b). 
In this case, sites can be classified into connected regions, and that feature 
is indicated in Fig.~\ref{fig:percolation}(c) using different shades. 
Each region is correlated and 
will correspond to a FM cluster. In this case all spins belong 
to some large cluster.
As the concentration $x$ grows, the clusters will tend to merge. 
As $x$ decreases,
these clusters will become more and more isolated.

Recent experimental work on (Ga,Cr)As have revealed unusual magnetic
properties which were associated with the random magnetism of the alloy.
The authors of Ref.~\onlinecite{dakhama} explained their results using a distributed
magnetic polaron model, that resembles the clustered-state ideas
discussed here and in Refs.~\onlinecite{paper1,timm,kaminski}. 

\section{Conclusions} \label{sec:conclusion}

In this paper, dynamical and transport properties of a single-band model
for diluted magnetic semiconductors have been presented. The calculations
were carried out on a lattice and using Monte Carlo techniques.
The optical conductivity, density of states, and resistance vs. temperature
agree with experimental data for Ga$_{1-x}$Mn$_x$As if the model is 
in a regime of {\it intermediate} $J/t$ coupling. In this region, the
carriers are neither totally localized at the Mn sites nor free, as 
in previous theories. The state of relevance has some
characteristics of a clustered state, in the sense that upon cooling 
from high temperatures first small regions are locally magnetized at a temperature
scale $T^*$ -- causing a mild insulating behavior -- while at an even
 lower temperature $T_{\rm C}$ the alignment of the individual cluster moments
occurs -- causing metallic behavior. This is in agreement with
previous investigations.\cite{paper1,paper0,timm,kaminski}

\begin{figure}[h]
\centering{
\includegraphics[width=6.5cm]{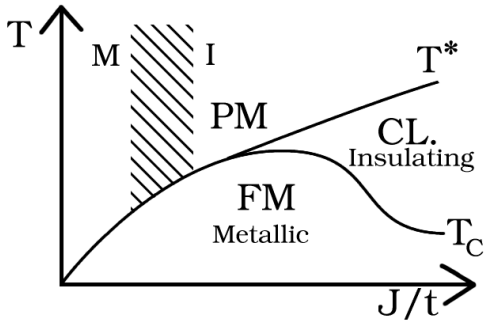}}
\caption{Sketch of the phase diagram indicating the conducting and insulating
regions, as obtained in the present investigations. The
dashed area  in the paramagnetic phase  
indicates the crossover region from a mild metallic weak coupling regime to
a mild insulating strong coupling regime.\label{fig:pd}}
\end{figure}

One of the main conclusions of the paper is summarized in Fig.\ref{fig:pd}
where the `transport' phase diagram is sketched. The low-temperature 
ferromagnetic phase is metallic (although often with poor metallicity).
The clustered state between $T^*$ and $T_{\rm C}$ has insulating properties,
and the same occurs in a good portion of the phase diagram above these two
characteristic temperatures. Note that here the terms metallic and insulating
simply refer to the slope of the resistivity vs. temperature curves. The
values of the resistivities in the two regimes are not very different, 
similarly as observed experimentally.

This and related efforts lead to a possible picture of DMS materials 
where the inhomogeneities play an important role. In this respect, 
these materials share characteristics with many other compounds such
as manganites and cuprates, where current  trends point toward the key
importance of nanocluster formation to understand the colossal
magnetoresistance and underdoped regions, respectively. 
``Clustered'' states appear to form a new paradigm that is useful
to understand the properties of many interesting materials.

\section{acknowledgments}

The authors acknowledge the help of J. A. Verg\'es in the
study of the conductance. The subroutines used in this context
were kindly provided by him.
Conversations with A. Fujimori are also gratefully acknowledged.
G. A. and E. D. are supported by the NSF grant DMR-0122523. Additional
 funds have been provided by Martech (FSU). 
Part of the computer simulations have been carried out on the CSIT IBM
SP3 system at the FSU School for Computational 
 Science and Information Technology (CSIT).


\begin{thebibliography}{99}
\bibitem{comment3} In Ref.~\onlinecite{schliemann}, a hybrid MC study of the multi-band model for DMS was
presented. In that reference the model was treated in k-space, approximating the fermionic
trace for its zero-temperature value. On the contrary, our present work treats the
fermionic trace exactly, and calculations are performed in real space.
\bibitem{comment1} The term `sketched' is
used instead of `established' due to the inevitable uncertainties of the
numerical effort, which is based on nearly exact results but obtained in small clusters.
\end{thebibliography}
\end{document}